\documentclass[conference,a4paper]{IEEEtran}

\usepackage{geometry}
\usepackage{graphicx}
\usepackage{color}
\usepackage{placeins}
\usepackage{float}
\usepackage{tabularx,colortbl}
\usepackage{amssymb}

\usepackage{amsthm}
\usepackage{cite}
\usepackage{amsmath}
\usepackage{caption2}

\usepackage{cases,subeqnarray}
\usepackage{bm,multirow,bigstrut}
\usepackage{textcomp}
\usepackage{latexsym,bm}
\usepackage{booktabs,changebar}
\usepackage{xcolor}
\usepackage{mathtools}
\usepackage{dsfont}
\usepackage{extarrows}
\usepackage{subfigure}

\usepackage{algorithm}
\usepackage{algpseudocode}

\theoremstyle{plain}
\newtheorem{thm}{Theorem}

\newtheorem{rem}{Remark}

\IEEEoverridecommandlockouts

\begin{document}

\title{Uplink Performance of RIS-aided Cell-Free Massive MIMO System Over Spatially Correlated Channels}
\vspace{-4mm}
\author{Enyu Shi, Jiayi~Zhang, Zhe Wang, Derrick Wing Kwan Ng,~\IEEEmembership{Fellow,~IEEE}, and Bo Ai,~\IEEEmembership{Fellow,~IEEE}
\thanks{E. Shi, J. Zhang, and Z. Wang are with the School of Electronics and Information Engineering, Beijing Jiaotong University, Beijing 100044, P. R. China. (e-mail: jiayizhang@bjtu.edu.cn).}
\thanks{D. W. K. Ng is with the School of Electrical Engineering and Telecommunications, University of New South Wales, NSW 2052, Australia. (e-mail: w.k.ng@unsw.edu.au).}
\thanks{B. Ai is with the State Key Laboratory of Rail Traffic Control and Safety, Beijing Jiaotong University, Beijing 100044, China.}}
\vspace{-4mm}

\maketitle

\begin{abstract}
We consider a practical spatially correlated reconfigurable intelligent surface (RIS)-aided cell-free (CF) massive multiple-input-multiple-output (mMIMO) system with multi-antenna access points (APs) over spatially correlated Rician fading channels. The minimum mean square error (MMSE) channel estimator is adopted to estimate the aggregated RIS channels. Then, we investigate the uplink spectral efficiency (SE) with the maximum ratio (MR) and the local minimum mean squared error (L-MMSE) combining at the APs and obtain the closed-form expression for characterizing the performance of the former. The accuracy of our derived analytical results has been verified by extensive Monte-Carlo simulations. Our results show that increasing the number of RIS elements is always beneficial, but with diminishing returns when the number of RIS elements is sufficiently large. Furthermore, the effect of the number of AP antennas on system performance is more pronounced under a small number of RIS elements, while the spatial correlation of RIS elements imposes a more severe negative impact on the system performance than that of the AP antennas.

\end{abstract}

\IEEEpeerreviewmaketitle

\section{Introduction}
Cell-free (CF) massive multiple-input-multiple-output (mMIMO) has recently been introduced as one of the most promising technologies for enabling beyond-fifth-generation (B5G) wireless communication systems \cite{zhang2020prospective}. In practice, cell-free systems are capable of serving a large number of user equipments (UEs) and improving the associated communication quality by shortening the distances between the UEs and the APs \cite{ngo2017cell,bjornson2019making}.
Meanwhile, reconfigurable intelligent surface (RIS) has recently been introduced as a new paradigm which can establish reconfigurable wireless channels/radio propagation environment \cite{wu2019intelligent}. In particular, the RIS can be deployed in scenarios with harsh communication environments, such as channels with large obstructions or interference to significantly improve communication quality \cite{wu2019intelligent,9743355,zhang2021ris}.

Recently, a number of studies have focused on how to jointly exploit the advantages of RIS and CF mMIMO to enhance the network performance.
For instance, in \cite{9743355}, the authors introduced an RIS-aided CF wireless energy transfer framework to improve the energy efficiency and the communication system quality.
Also, in \cite{zhang2020reconfigurable}, the authors deployed multiple RISs to assist CF massive MIMO and proposed an iterative resource allocation algorithm to address the system sum-rate optimization problem.
Besides, in \cite{bashar2020performance}, the authors proposed a new channel estimation method for RIS-aided CF communications, which can estimate the involved end-to-end channels by switching each element in turn.
However, none of the aforementioned studies considered the effect of spatial correlation of RIS on system performance.
Indeed, spatial correlations among RIS elements naturally exist in practice due to their sub-wavelength structure. As such, in \cite{bjornson2020rayleigh}, a sinc function-based model was proposed to capture the spatial correlations that are determined by the spacing among RIS elements.
Based on this, in \cite{van2021reconfigurable}, the authors proposed an aggregated channel estimation method and considered the performance of a CF system assisted by a single RIS with spatially correlated elements.
However, only Rayleigh fading channels with single-antenna APs were considered and their results are not applicable to the practical RIS-aided CF mMIMO systems. Therefore, there is an emerging need to consider the impact of the correlations of the RIS elements and AP antennas to study the system performance limit.

To address the above limitations, we consider an RIS-aided CF mMIMO system with multi-antenna APs over spatially correlated Rician fading channels. More specifically, we propose the minimum mean square error (MMSE)-based channel estimation to estimate the aggregated channels between the APs and the UEs. Then, we utilize the maximum ratio (MR) and the local minimum mean squared error (L-MMSE) combining at the APs to obtain the uplink spectral efficiency (SE) and obtain the closed-form expressions for that adopting MR combining. The results show that increasing the number of RIS elements or AP antennas can always improve the system performance. Moreover, the spatial correlation among RIS elements dominates the system performance more than that of AP antennas. Finally, compared with MR combining, L-MMSE combining enjoys a better performance with a large number of RIS elements at the cost of higher computational complexity.

\textbf{Notation:} The superscripts $\mathbf{x}^{H}$ and $x^\mathrm{*}$ are used to represent conjugate transpose and conjugate, respectively.
The matrices and column vectors are denoted by boldface uppercase letters $\mathbf{X}$ and boldface lowercase letters $\mathbf{x}$, respectively.
The ${\rm{mod}}\left( { \cdot , \cdot } \right)$, $\left\|  \cdot  \right\|$, and $\left\lfloor  \cdot  \right\rfloor $ denote the modulus operation, the Euclidean norm and the truncated argument, respectively. ${\rm{tr}}\left(  \cdot  \right)$, $\mathbb{E}\left\{  \cdot  \right\}$, and ${\rm{Cov}}\left\{  \cdot  \right\}$ are the trace, expectation and covariance operators. $\otimes$ denotes the Kronecker products.
Finally, the circularly symmetric complex Gaussian random variable $x$ with variance $\sigma^2$ is denoted by $x \sim \mathcal{C}\mathcal{N}\left( {0,{\sigma^2}} \right)$.

\section{System Model}\label{se:model}
As shown in Figure 1, we consider an RIS-aided CF mMIMO system consisting of $M$ APs, one RIS, and $K$ UEs. The RIS has $N$ reflective elements that can introduce some phase shifts to the incident signals. We consider that each AP is equipped with $L$ antennas and each UE is equipped with a single antenna. The central processing unit (CPU) connect to all APs via fronthaul links. The standard time division duplex (TDD) protocol is adopted, where $\tau_c$ is the length of each coherence block. We assume that $\tau_p$ symbols are exploited for the channel estimation phase in the uplink (UL) and $\tau_u = \tau_c - \tau_p$ symbols are utilized for data transmission.

Let ${{\bf{g}}_{mk}}\in {\mathbb{C}}{^{L}}$ denote the direct link channel between UE $k$ and AP $m$. ${{\bf{H}}_m} \in{\mathbb{C}} {^{N \times L}}$ denote the channel matrix from AP $m$ to the RIS and ${{\bf{z}}_k} \in {\mathbb{C}}{^{N}}$ denote the channel from UE $k$ to the RIS, respectively. We consider a realistic model to capture the spatial correlation among the RIS elements \cite{bjornson2020rayleigh}. Meanwhile, the spatial correlation due to the AP multiple antennas is also considered in this paper. We assume that AP-UE channels are Rayleigh fading whereas RIS-UE and AP-RIS channels are Rician fading. Then, the channels ${{\bf{g}}_{mk}}$, ${{\bf{z}}_k}$, and ${{\bf{H}}_m}$ can be modeled as
\begin{align}
{{\bf{g}}_{mk}} &\sim {\cal C}{\cal N}\left( {0,{{\mathbf{R}}_{mk}}} \right),\\
{{\bf{H}}_m} &= {{{\bf{\bar H}}}_m} + {{{\bf{\tilde H}}}_m},\;{{\bf{z}}_k} = {{\bf{\Theta }}_k}{{{\bf{\bar z}}}_k} + {{{\bf{\tilde z}}}_k},
\end{align}
where ${\mathbf{R}}_{mk}\in {\mathbb{C}}{^{L \times L}}$ is the spatial correlation matrix of AP antennas and ${\beta _{mk}} = {{{\rm{tr}}\left( {{{\bf{R}}_{mk}}} \right)} \mathord{\left/
 {\vphantom {{{\rm{tr}}\left( {{{\bf{R}}_{mk}}} \right)} L}} \right.
 \kern-\nulldelimiterspace} L}$ is the large-scale fading coefficient between AP $m$ and UE $k$. ${{{\bf{\bar H}}}_m} \in {{\mathbb{C}}^{N \times L}}$ and ${{{\bf{\bar z}}}_k} \in {{\mathbb{C}}^{N}}$ represent the deterministic LoS component of AP-RIS and RIS-UE channels, respectively. ${{{\bf{\tilde H}}}_m} \sim {\cal C}{\cal N}\left( {{\bf{0}},{{{\bf{\tilde R}}}_m}} \right)$ and ${{{\bf{\tilde z}}}_k} \sim {\cal C}{\cal N}\left( {{\bf{0}},{{{\bf{\tilde R}}}_k}} \right)$ are the NLoS components, where ${{{\bf{\tilde R}}}_m} = \frac{1}{{LN{\beta _m}}}\left( {{\bf{R}}_m^T \otimes {{\bf{R}}_r}} \right)\in {{\mathbb{C}}^{NL \times NL}}$ and ${{{\bf{\tilde R}}}_k} = {\beta _k}{\bf{R}} \in {{\mathbb{C}}^{N \times N}}$ \cite{zhang2021improving,wang04962}. ${{\beta _m}}$ and ${{\beta _k}}$ denote the large-scale fading coefficients from AP $m$ and UE $k$ to RIS, respectively. Note that ${{\bf{R}}_m}$ and ${{{\bf{R}}_r} = {\beta_m}\bf{R}}$ denote the correlation matrices of the AP side and RIS side, respectively. ${\mathbf{R}} \in {{\mathbb{C}}^{N \times N}}$ characterizes the spatial correlation of the RIS which has the $\left( {m',n'} \right)$-th element as ${{\left[ \mathbf{R} \right]_{m'n'}} = {\rm{sinc}}\left( {{{2\left\| {{{\mathbf{u}}_{m'}} - {{\mathbf{u}}_{n'}}} \right\|} \mathord{\left/
 {\vphantom {{2\left\| {{{\mathbf{u}}_{m'}} - {{\mathbf{u}}_{n'}}} \right\|} \lambda }} \right.
 \kern-\nulldelimiterspace} \lambda }} \right)}$, where ${{\rm{sinc}}\left( x \right) = {{\sin \left( {\pi x} \right)} \mathord{\left/
 {\vphantom {{\sin \left( {\pi x} \right)} {\left( {\pi x} \right)}}} \right.
 \kern-\nulldelimiterspace} {\left( {\pi x} \right)}}}$ denotes the sinc function and ${\lambda}$ denotes the carrier wavelength \cite{bjornson2020rayleigh}.
Besides, ${{\mathbf{u}_x} \!=\! {\left[ {0,\bmod \left( {x - 1,{N_H}} \right){d_H},\left\lfloor {{{\left( {x - 1} \right)} \mathord{\left/
 {\vphantom {{\left( {x - 1} \right)} {{N_H}}}} \right.
 \kern-\nulldelimiterspace} {{N_H}}}} \right\rfloor {d_V}} \right]^T}}$, $x \!\in\! \left\{ {m',n'} \right\}$ is the position vector, where ${N_{V}}$ and ${N_{H}}$ are the number of elements at RIS in each column and row, respectively, such that ${N = {N_H} \times {N_V}}$.
Moreover, ${{\bf{\Theta }}_k} = {\rm{diag}}\left( {{e^{j{\theta _{1k}}}}, \cdots ,{e^{j{\theta _{Nk}}}}} \right) \in {{\mathbb{C}}^{N \times N}}$, where ${\theta _{nk}} \in \left[ { - \pi ,\pi } \right]$ is the phase-shift of the LoS component between UE $k$ and the $n$-th element of RIS caused by the small change in the UE location \cite{ozdogan2019performance}. We assume that all elements of ${{\bf{\Theta }}_k}$ are idential \cite{ozdogan2019performance} such that the LoS component can be written as ${{{\bf{\bar z}}}_k}{e^{j{\theta _k}}}$.

Let ${\mathbf{\Phi}  = {\rm{diag}}\left( {{e^{j{\varphi _1}}},{e^{j{\varphi _2}}}, \cdots ,{e^{j{\varphi _N}}}} \right)}$ denote the phase shift matrix of RIS, where ${\varphi _n} \in \left[ { - \pi ,\pi } \right],\forall n \in \left\{ {1, \ldots ,N} \right\}$, denotes the phase shift introduced by the $n$-th RIS element. Thus, the total uplink aggregated channel between UE $k$ and AP $m$ can be formulated as
\begin{align}
{{\bf{o}}_{mk}} = {{\bf{g}}_{mk}} + {\bf{H}}_m^H{\bf{\Phi }}{{\bf{z}}_k},\forall m,k,
\end{align}
which consists of a direct link between AP $m$ and UE $k$ and a cascaded link reflected by the RIS. By mathematical derivation, the end-to-end channel is obtained as
\begin{align}
&{{\bf{o}}_{mk}} = {{\bf{g}}_{mk}} + {\bf{H}}_m^H{\bf{\Phi }}{{\bf{z}}_k}\notag\\
 &= {{\bf{g}}_{mk}} + {\left( {{{{\bf{\bar H}}}_m} + {{{\bf{\tilde H}}}_m}} \right)^H}{\bf{\Phi }}\left( {{{{\bf{\bar z}}}_k}{{e^{j{\theta _k}}}} + {{{\bf{\tilde z}}}_k}} \right)\notag\\
 &=\! \underbrace {{\bf{\bar H}}_m^H{\bf{\Phi }}{{{\bf{\bar z}}}_k}}_{{{{\bf{\bar o}}}_{mk}}}{e^{j{\theta _k}}} \!+\! \underbrace {{{\bf{g}}_{mk}} \!+\! {\bf{\bar H}}_m^H{\bf{\Phi }}{{{\bf{\tilde z}}}_k} \!+\! {\bf{\tilde H}}_m^H{\bf{\Phi }}{{{\bf{\bar z}}}_k}{e^{j{\theta _k}}} \!\!+\! {\bf{\tilde H}}_m^H{\bf{\Phi }}{{{\bf{\tilde z}}}_k}}_{{{{\bf{\tilde o}}}_{mk}}}.
\end{align}
We consider that UEs move slowly, so the LoS component ${{{{\bf{\bar o}}}_{mk}}}$ is slow time-varying and is known at the APs \cite{ozdogan2019performance,zheng2022cell}. In the following, we first analyze the second-order statistic of unknown ${{\bf{\tilde o}}_{mk}}$ that will be handy for the analysis in the sequel.
\begin{thm}
The covariance matrix of the second term ${{{{\bf{\tilde o}}}_{mk}}}$ in (4) can be obtained as
\begin{align}
\mathbb{E}\left\{ {{ {{{\bf{\tilde o}}}_{mk}}{{{\bf{\tilde o}}}^{H}_{mk}} }} \right\} \!=\! \underbrace {{{\bf{R}}_{mk}} \!+\! {\bf{\bar H}}_m^H\mathbf{\Phi} {{{\bf{\tilde R}}}_k}{\mathbf{\Phi} ^H}{{{\bf{\bar H}}}_m} \!+\! {\bf{Q}}_{mk}^1 \!+\! {\bf{Q}}_{mk}^2}_{{\bf{R}}_{mk}^o}.
\end{align}
\end{thm}
\begin{IEEEproof}
The proof is given in Appendix A.
\end{IEEEproof}
\begin{rem}
Note that the mean and covariance matrix of the aggregated channel ${{{{\bf{o}}}_{mk}}}$ can be obtained at AP in each coherence block, so we can utilize them to estimate the end-to-end aggregated channel in the next subsection.
\end{rem}

\begin{figure}[t]\vspace{0.1in}
\centering
\includegraphics[scale=0.4]{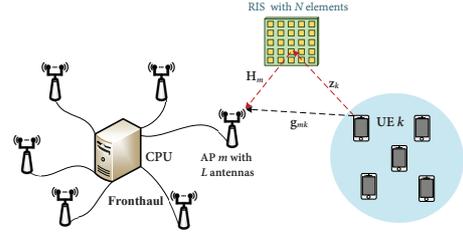}
\caption{RIS-aided CF mMIMO system with multi-antenna APs.
\label{Figure_1}}\vspace{-5mm}
\end{figure}
\subsection{Uplink Channel Estimation}
We adopt $\tau_p$ pilot sequences which are mutually orthogonal for channel estimation in each coherence block. All the UEs share the same $\tau_p$ orthogonal pilot sequences. In particular, the pilot sequence of UE $k$ is denoted by ${\bm{\phi} _k} \in {{\mathbb{C}}^{{\tau _p}}}$ and satisfies ${{\left\| {{\bm{\phi} _k}} \right\|^2} = {\tau_p}}$.
Let ${{\cal P}_k}$ denotes the index subset of UEs which adopts the same pilot sequence as UE $k$ including itself because of $K > {\tau _p}$. The received pilot signals ${\bf{Y}}_m^p \in {\mathbb{C}^{L \times {\tau _p}}}$ from all the UEs at AP $m$ as
\begin{align}
{\bf{Y}}_m^p = \sum\limits_{k = 1}^K {\sqrt {{{\hat p}_k}} {{\bf{o}}_{mk}}} {\bm{\phi}} _k^T + {\bf{N}}_m^p,
\end{align}
where ${{{\hat p}_k}}$ is the pilot transmit power of UE $k$, ${\bf{N}}_m^p \in {\mathbb{C}^{L \times {\tau _p}}}$ is the additive noise with independent ${\cal C}{\cal N}\left( {0,{\sigma ^2}} \right)$ entries, and ${\sigma ^2}$ is the noise power. We multiply the received signal by ${{\bm{\phi}} _k^ *}$ at AP $m$ to estimate ${{{\bf{o}}_{mk}}}$ and the results can be obtained as
\begin{align}
{\bf{y}}_m^p \!=\! {\bf{Y}}_m^p{\bm{\phi}} _k^ *  \!=\! \sqrt {{{\hat p}_k}} {\tau _p}{{\bf{o}}_{mk}} \!+\!\! \sum\limits_{i \in {{\cal P}_k}\backslash \left\{ k \right\}} \!\!{\sqrt {{{\hat p}_i}} {\tau _p}{{\bf{o}}_{mi}}}  \!+\! {\bf{n}}_m^p,
\end{align}
where ${\bf{n}}_m^p = {\bf{N}}_m^p{\bm{\phi}} _k^ *  \sim {\cal C}{\cal N}\left( {0,{\sigma ^2}{{\bf{I}}_L}} \right)$. Based on (7), if ${{{{\bf{\bar o}}}_{mk}}}$, ${\bf{R}}_{mk}^o$, and ${{\varphi _k}}$ are available at AP $m$, we utilize the phase-aware MMSE channel estimator to estimate the effective channel ${{\bf{o}}_{mk}}$ as
\begin{align}
{{{\bf{\hat o}}}_{mk}} = {{{\bf{\bar o}}}_{mk}}{e^{j{\varphi _k}}} + \sqrt {{\hat p_k}} {\bf{R}}_{mk}^o{\bf{\Psi }}_{mk}^{ - 1}\left( {{\bf{y}}_{mk}^p - {\bf{\bar y}}_{mk}^p} \right),
\end{align}
where ${\bf{\bar y}}_{mk}^p = \sum\nolimits_{i \in {{\cal P}_k}} {\sqrt {{{\hat p}_i}} } {\tau _p}{{{\bf{\bar o}}}_{mi}}{e^{j{\varphi _i}}}$ and ${{\bf{\Psi }}_{mk}} = \sum\nolimits_{i \in {{\cal P}_k}} {{{\hat p}_i}{\tau _p}{\bf{R}}_{mi}^o}  + {\sigma ^2}{{\bf{I}}_L}$. The estimated channel ${{{\bf{\hat o}}}_{mk}}$ and the estimation error ${{{\bf{\tilde o}}}_{mk}} = {{\bf{o}}_{mk}} - {{{\bf{\hat o}}}_{mk}}$ are independent random variable with
\begin{align}
&\mathbb{E}\left\{ {{{{\bf{\hat o}}}_{mk}}\left| {{\varphi _k}} \right.} \right\} = {{{\bf{\bar o}}}_{mk}}{e^{j{\varphi _k}}},{\rm{Cov}}\left\{ {{{{\bf{\hat o}}}_{mk}}\left| {{\varphi _k}} \right.} \right\} = {{\hat p}_k}{\tau _p}{{\bf{\Omega }}_{mk}},\notag\\
&\mathbb{E}\left\{ {{{{\bf{\tilde o}}}_{mk}}} \right\} = {\bf{0}},\quad \quad \quad \quad \;\;\;{\rm{Cov}}\left\{ {{{{\bf{\tilde o}}}_{mk}}} \right\} = {{\bf{C}}_{mk}},\notag
\end{align}
where ${{\bf{\Omega }}_{mk}} = {\bf{R}}_{mk}^o{\bf{\Psi }}_{mk}^{ - 1}{\bf{R}}_{mk}^o$ and ${{\bf{C}}_{mk}} = {\bf{R}}_{mk}^o - {{\hat p}_k}{\tau _p}{\bf{R}}_{mk}^o{\bf{\Psi }}_{mk}^{ - 1}{\bf{R}}_{mk}^o$.

\subsection{Uplink Data Transmission}
In the uplink, we consider that the UEs transmit the uplink data to all the APs simultaneously. Then, the received signal at AP $m$ is obtained as
\begin{align}
{{\bf{y}}_m} = \sum\limits_{k = 1}^K {{{\bf{o}}_{mk}}} {s_k} + {{\bf{n}}_m},
\end{align}
where ${s_k} \sim {\cal C}{\cal N}\left( {0,{p_k}} \right)$ denotes the uplink signal transmitted by UE $k$ with power ${p_k} = \mathbb{E}\left\{ {{{\left| {{s_k}} \right|}^2}} \right\}$. ${{\bf{n}}_m} \sim {\cal C}{\cal N}\left( {{\bf{0}},{\sigma ^2}{{\bf{I}}_L}} \right)$ denotes the additive noise. We consider that each AP can process the uplink data locally and intermediately with a combining vector. We adopt ${{\bf{v}}_{mk}} \in {\mathbb{C}^L} $ denoting the combining vector which is designed by AP $m$ for UE $k$ and then AP $m$ can obtain the local estimate of ${s_k}$ as
\begin{align}
{{\tilde s}_{mk}} = {\bf{v}}_{mk}^H{{\bf{y}}_m} = \sum\limits_{k = 1}^K {{\bf{v}}_{mk}^H{{\bf{o}}_{mk}}} {s_k} + {\bf{v}}_{mk}^H{{\bf{n}}_m}.
\end{align}
Any combining vector is available for (10) and the local channel state information (CSI) in AP $m$ can be used to design ${{\bf{v}}_{mk}}$. Here, we consider the MR combining with ${{\bf{v}}_{mk}}= {\bf{\hat o}}_{mk}$ and the L-MMSE combining introduced in \cite{wang2020uplink} as
\begin{align}
{{\bf{v}}_{mk}} \!=\! {p_k}{\left( {\sum\limits_{i = 1}^K {{p_i}\!\left( {{{{\bf{\hat o}}}_{mi}}{{\left( {{{{\bf{\hat o}}}_{mi}}} \right)}^H} \!\!+\! {{\bf{C}}_{mi}}} \right)}  \!+\! {\sigma ^2}{{\bf{I}}_L}} \right)^{ - 1}}\!\!{{{\bf{\hat o}}}_{mk}}.
\end{align}
Note that (11) can minimize ${\rm{MS}}{{\rm{E}}_{mk}} = \mathbb{E}\left\{ {\left. {{{\left| {{s_k} - {\bf{v}}_{mk}^H{{\bf{y}}_m}} \right|}^2}} \right|\left\{ {{{{\bf{\hat o}}}_{mk}}} \right\}} \right\}$ and is optimal for the MMSE estimation.

After the MR/L-MMSE combining, all APs convey the local estimates ${{\tilde s}_{mk}}$ in (10) to the CPU. To further mitigate the inter-user interference \cite{bjornson2019making}, ${{\tilde s}_{mk}}$ are linearly weighted with the large-scale fading decoding (LSFD) coefficients ${a_{mk}^ * }\in \mathbb{C}$ as
\begin{align}
{{\hat s}_k} = \sum\limits_{m = 1}^M {a_{mk}^ * } {{\tilde s}_{mk}} = {\bf{a}}_k^H{{\bf{u}}_{kk}}{s_k} + \sum\limits_{i \ne k}^K {{\bf{a}}_k^H{{\bf{u}}_{ki}}{s_i}}  + {{\bf{n}}_k},
\end{align}
where ${{\bf{u}}_{ki}} = {\left[ {{\bf{v}}_{1k}^H{{\bf{o}}_{1i}}, \cdots ,{\bf{v}}_{Mk}^H{{\bf{o}}_{Mi}}} \right]^T} \in \mathbb{C}{^M}$, ${{\bf{n}}_k} = \sum\nolimits_{m = 1}^M {a_{mk}^ * {\bf{v}}_{mk}^H{{\bf{n}}_m}} $, and ${{\bf{a}}_k} = {\left[ {{a_{1k}}, \cdots ,{a_{Mk}}} \right]^T} \in \mathbb{C} {^M}$ denotes the LSFD coefficient vector, respectively.

\section{Performance Analysis}\label{se:performance}
In this section, we analyze the uplink performance of the RIS-aided CF mMIMO system with the MR combining scheme and investigate the impact of spatial correlation. Based on (12), the uplink achievable SE lower bound of UE $k$ can be obtained by utilizing the use-and-then-forget (UatF) bound \cite{bjornson2019making,zhang2021local} as follows
\begin{align}
{\rm{S}}{{\rm{E}}_k} = \frac{{{\tau _u}}}{{{\tau _c}}}{\log _2}\left( {1 + {\gamma _k}} \right),
\end{align}
with the effective signal-to-interference-plus-noise ratio (SINR) ${\gamma _k}$ given by
\begin{align}
\!{\gamma _k} \!=\! \frac{{{p_k}{{\left| {{\bf{a}}_k^H\mathbb{E}\left\{ {{{\bf{u}}_{kk}}} \right\}} \right|}^2}}}{{{\bf{a}}_k^H\!\!\left( {\sum\limits_{i = 1}^K {{p_i}{{\bf{T }}_{ki}}} \!-\! {p_k}{ {\mathbb{E}\left\{ {{{\bf{u}}_{kk}}} \right\}}{\mathbb{E}\!\left\{ {{{\bf{u}}^{H}_{kk}}} \right\}} } \!+\! {\sigma ^2}{{\bf{D}}_k}} \right)\!{{\bf{a}}_k}}},
\end{align}
where ${{\bf{T }}_{ki}} = \left[ \mathbb{E}{\left\{ {{\bf{v}}_{mk}^H{{\bf{o}}_{mi}}{\bf{o}}_{m'i}^H{{{\bf{v}}}_{m'k}}} \right\}:\forall m,m'} \right] \in {\mathbb{C}^{M \times M}}$,
${{\bf{D}}_k} = {\rm{diag}}\left( {\mathbb{E}\left\{ {{{\left\| {{{{\bf{v}}}_{1k}}} \right\|}^2}} \right\}, \cdots ,\mathbb{E}\left\{ {{{\left\| {{{{\bf{v}}}_{Mk}}} \right\|}^2}} \right\}} \right) \in \mathbb{C}{^{M \times M}}$. To maximize the effective SINR in (13), the CPU can optimize ${{{\bf{a}}_k}}$ from \cite{wang2020uplink} as
\begin{align}
\!\!\!{{\bf{a}}_k} \!\!=\!\! {\left( \!{\sum\limits_{i = 1}^K {{p_i}{{\bf{T }}_{ki}}}  \!-\! {p_k}{ {\mathbb{E}\!\left\{ {{{\bf{u}}_{kk}}} \right\}}{\mathbb{E}\!\left\{ {{{\bf{u}}^{H}_{kk}}} \right\}} } \!+\! {\sigma ^2}{{\bf{D}}_k}} \right)^{ - 1}}\!\!\!\!\!\!\mathbb{E}\!\left\{ {{{\bf{u}}_{kk}}} \!\right\}.
\end{align}
The closed-form expressions of the SE cannot be obtained when we adopt the MMSE combining, here, we focus on the MR combining to obtain the closed-form SE expressions with ${{\bf{v}}_{mk}} = {{{\bf{\hat o}}}_{mk}}$.

\newcounter{mytempeqncnt}
\begin{figure*}[t!]\vspace{0.1in}
\normalsize
\setcounter{mytempeqncnt}{1}
\setcounter{equation}{15}
\begin{align}\label{SINR_CF}
{\gamma _k} = \frac{{{p_k}{{\left| {{\rm{tr}}\left( {{\bf{A}}_k^H{{\bf{Z}}_k}} \right)} \right|}^2}}}{{\sum\limits_{i = 1}^K {{p_i}{\rm{tr}}\left( {{\bf{A}}_k^H{{\bf{\Xi }}_{ki}}{{\bf{A}}_k}} \right)}  + \sum\limits_{i \in {{\cal P}_k}\backslash \left\{ k \right\}} {{p_i}{{{\Gamma }}_{ki}}}  + {\rm{tr}}\left( {{\bf{A}}_k^H\left( {{\sigma ^2}{{\bf{Z}}_k} - {p_k}{\bf{J}}_k^2} \right){{\bf{A}}_k}} \right)}}.
\end{align}
\setcounter{equation}{16}
\hrulefill
\end{figure*}

\begin{thm}
The closed-form expression for the uplink SE of UE $k$ is given by (13), where the SINR is expressed as (16) at the top of the next page. Then, the desired signal can be expressed by the following formula
\begin{align}
{{\bf{A}}_k} = {\rm{diag}}\left( {{a_{1k}}, \cdots ,{a_{Mk}}} \right) \in {\mathbb{C}^{M \times M}},\\
{{\bf{Z}}_k} = {\rm{diag}}\left( {{z_{1k}}, \cdots ,{z_{Mk}}} \right) \in {\mathbb{C}^{M \times M}},\\
{z_{mk}} = {\rm{tr}}\left( {{p_k}{\tau _p}{{\bf{\Omega }}_{mk}} + {{{\bf{\bar o}}}_{mk}}{{\left( {{{{\bf{\bar o}}}_{mk}}} \right)}^H}} \right).
\end{align}
Also, the definition of non-coherent interference ${{{\bf{\xi }}_{ki}}}$ is shown as
\begin{align}
{{\bf{\Xi }}_{ki}} = {\rm{diag}}\left( {{\xi _{1,ki}}, \cdots ,{\xi _{M,ki}}} \right) \in {\mathbb{C}^{M \times M}},
\end{align}
\begin{align}
{\xi _{m,ki}} &= {\hat p_k}{\tau _p}{\rm{tr}}\left( {{\bf{R}}_{mi}^o{{\bf{\Omega }}_{mk}}} \right) + {\bf{\bar o}}_{mk}^H{\bf{R}}_{mi}^o{{{\bf{\bar o}}}_{mk}} \notag\\
&+ {\hat p_k}{\tau _p}{\bf{\bar o}}_{mi}^H{{\bf{\Omega }}_{mk}}{{{\bf{\bar o}}}_{mi}} + {\left| {{\bf{\bar o}}_{mk}^H{{{\bf{\bar o}}}_{mi}}} \right|^2}.
\end{align}
The coherent interference ${{{\bf{\Gamma }}_{ki}}}$ is shown as
\begin{align}
&{{{\Gamma }}_{ki}} = {\hat p_k}{\hat p_i}\tau _p^2{\left| {{\rm{tr}}\left( {{{\bf{A}}_k}{{\bf{\Delta }}_{ki}}} \right)} \right|^2},\\
&{{\bf{\Delta }}_{ki}} = {\rm{diag}}\left( {{\varpi _{1,ki}}, \cdots ,{\varpi _{M,ki}}} \right) \in {{\mathbb{C}}^{M \times M}},\\
&{\varpi _{m,ki}}{\rm{ = tr}}\left( {{\bf{R}}_{mi}^o{\bf{\Psi }}_{mk}^{ - 1}{\bf{R}}_{mk}^o} \right).
\end{align}
Finally, ${{\bf{J}}_k} = {\rm{diag}}\left( {{{\left\| {{{{\bf{\bar o}}}_{1k}}} \right\|}^2}, \cdots ,{{\left\| {{{{\bf{\bar o}}}_{Mk}}} \right\|}^2}} \right)$.
\end{thm}
\begin{IEEEproof}
The proof is given in Appendix B.
\end{IEEEproof}
\begin{rem}
Note that the spatial correlation and the aggregated channels via the RIS affect the system performance by affecting ${{\bf{R}}_{mk}^o}$ in the closed-form (16). By reducing the spatial correlation of RIS elements or increasing the number of RIS elements, the increase of ${{\bf{R}}_{mk}^o}$ leads to the increase of the numerator in (16), and the system performance is improved.
\end{rem}
\begin{rem}
In CF mMIMO systems, the spatial correlation of the AP antennas is beneficial to the SE \cite{wang2020uplink}, while in our system, the aggregated channel consists the interaction between ${{\bf{R}}_m}$ and ${{\bf{R}}}$. As such the conclusion in \cite{wang2020uplink} may no longer hold, especially when the RIS elements are strongly correlated.
\end{rem}

\section{Numerical Results and Discussion}\label{se:numerical}
We provide some numerical results to verify the accuracy of the derived analysis and evaluate the performance of the RIS-aided CF mMIMO system over spatial correlation of RIS elements and AP antennas. We assume that the APs and UEs are uniformly distributed in the $1 \times 1\,{\rm{k}}{{\rm{m}}^2}$ and $0.1 \times 0.1\,{\rm{k}}{{\rm{m}}^2}$ area with a wrap-around scheme \cite{bjornson2019making}, respectively. Moreover, the RIS is located at the regional center. The height of the APs, UEs, and RIS is $15$ m, $1.65$ m, and $30$ m, respectively. For the pathloss, we take AP-RIS as an example which consists of a LoS path and we utilize the COST 321 Walfish-Ikegami model \cite{wang04962} to compute the pathloss as
\begin{align}
{\beta _m}\left[ {{\rm{dB}}} \right] =  - 30.18 - 26{\log _{10}}\left( {\frac{{{d_m}}}{{1{\rm{m}}}}} \right) + {F_m},
\end{align}
where $d_m$ denotes the distance between AP $m$ and RIS. The Rician $\kappa$-factor is denoted as ${\kappa _m} = {10^{1.3 - 0.003{d_m}}}$.
The shadow fading $F_{m}$ and other parameters is similar to \cite{ngo2017cell}. The large-scale coefficients of ${{\bf{H}}_m}$ are given by
\begin{align}
\beta _m^{{\rm{LoS}}} = \frac{{{\kappa _m}}}{{{\kappa _m} + 1}}{\beta _m},{\kern 1pt} \;\;\beta _m^{{\rm{NLoS}}} = \frac{1}{{{\kappa _m} + 1}}{\beta _m}.
\end{align}
The Gaussian local scattering model in \cite{bjornson2017massive} is utilized to generate the spatial correlation matrix ${{\bf{R}}_{mk}}$. The ${{\bf{R}}_{mk}}$ $\left( {l,n} \right)$-th element can be written as
\begin{align}
{\left[ {{{\bf{R}}_{mk}}} \right]_{ln}} = \frac{{\beta _{mk}^{{\rm{NLoS}}}}}{{\sqrt {2\pi } {\sigma _\varphi }}}\int_{ - \infty }^{ + \infty } {{e^{j2\pi {d_H}\left( {l - n} \right)\sin \left( {{\theta _{mk}} + \delta } \right)}}} {e^{ - \frac{{{\delta ^2}}}{{2\sigma _\varphi ^2}}}}d\delta ,\notag
\end{align}
where $\delta  \sim {\cal N}\left( {0,\sigma _\varphi ^2} \right)$ is the distributed deviation from ${\theta_{mk}}$ with angular standard deviation (ASD) ${{\sigma _\varphi }}$. Every UE transmits with power $23$ dBm and the noise power  ${\sigma ^2} = ?94$ dBm. As for the phase shift design of RIS, we take a fixed value that the $N$ elements phase shift is set equal to ${\pi  \mathord{\left/
 {\vphantom {\pi  4}} \right.
 \kern-\nulldelimiterspace} 4}$ \cite{van2021reconfigurable}.

\begin{figure}[t]
\centering
\includegraphics[scale=0.4]{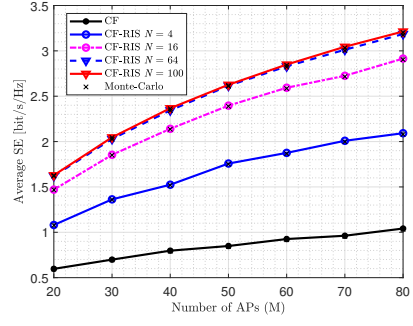}
\caption{Average SE per UE versus different number of APs and RIS elements with the MR combining (${K = 10}$, ${L = 1}$, ${\tau_p = 5}$, ${{d_{\rm{V}}}={d_{\rm{H}}} = \frac{1}{2}\lambda }$). \vspace{-4mm}
\label{Figure_2}}
\end{figure}
\begin{figure}[t]
\centering
\includegraphics[scale=0.4]{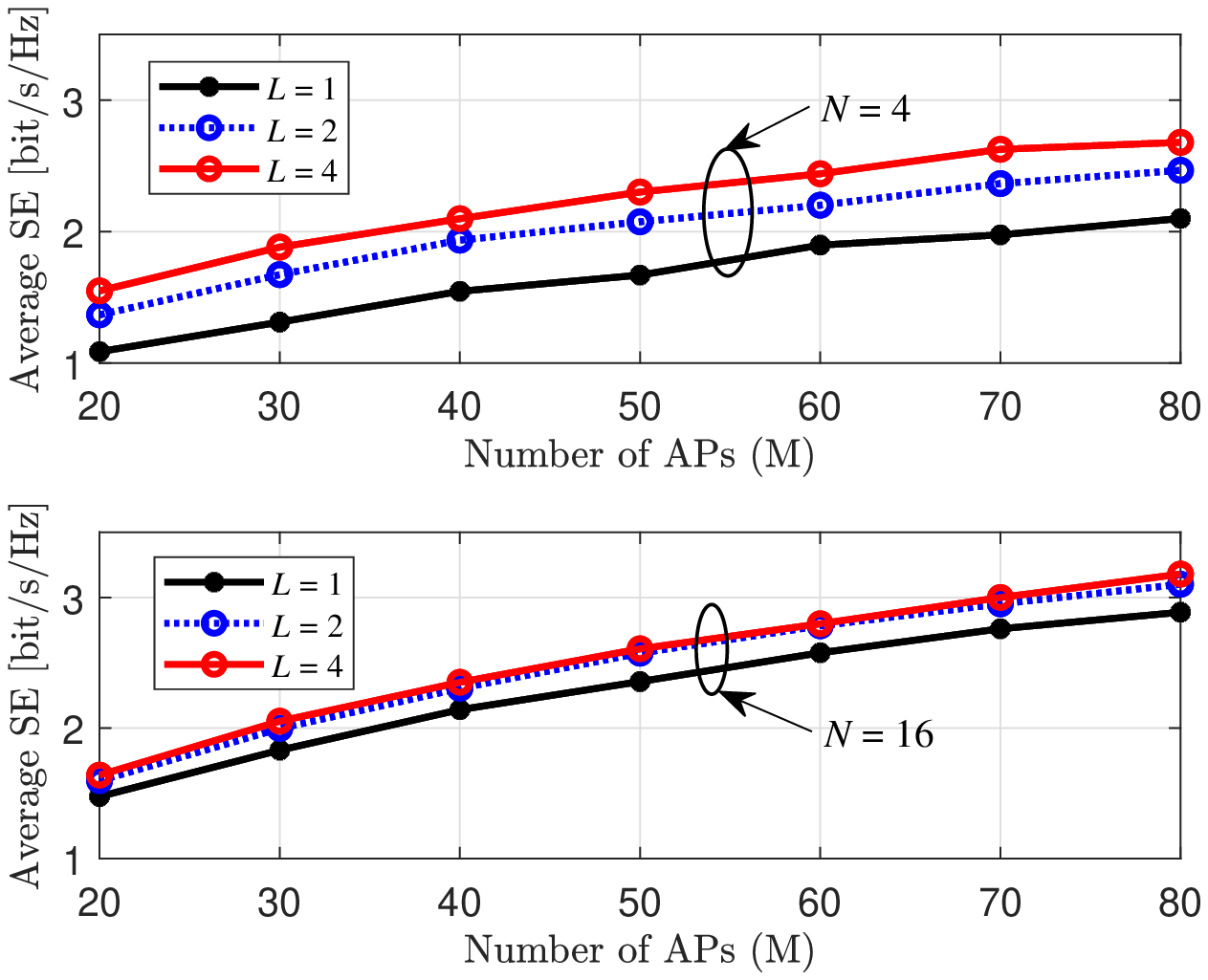}
\caption{Average SE per UE versus different number of AP antennas and RIS elements with the MR combining (${K = 10}$, ${\tau_p = 5}$, ${{d_{\rm{V}}}={d_{\rm{H}}} = \frac{1}{2}\lambda }$). \vspace{-4mm}
\label{Figure_3}}
\end{figure}

\begin{figure}[t]
\centering
\includegraphics[scale=0.4]{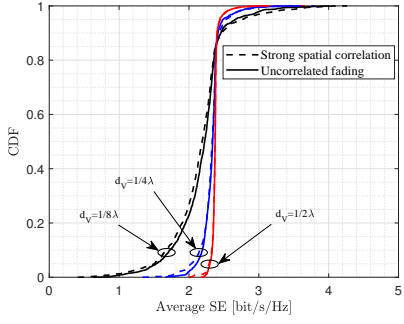}
\caption{CDF of the uplink average SE per UE under different spatial correlations of AP antennas and RIS elements (${M = 40}$, ${K = 10}$, ${N = 36}$, ${\tau_p = 5}$).\vspace{-4mm}
\label{Figure_4}}
\end{figure}
\begin{figure}[t]
\centering
\includegraphics[scale=0.4]{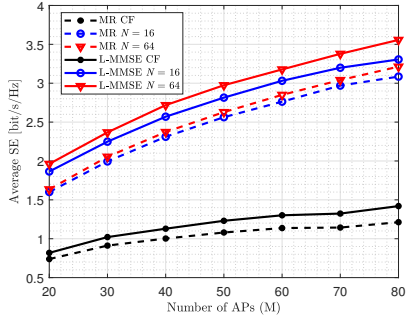}
\caption{Average SE for per UE against different numbers of RIS elements with the MR and the L-MMSE combining (${L = 2}$, ${K = 10}$, ${N = 16, 64}$, ${\tau_p = 5}$, ${{d_{\rm{V}}}={d_{\rm{H}}} = \frac{1}{2}\lambda }$).\vspace{-4mm}
\label{Figure_5}}
\end{figure}
Figure~\ref{Figure_2} shows the average SE per UE as a function of the number of RIS elements $N$. It is clear that in the considered setting, the per UE uplink SE increases with the number of RIS $N$ and AP $M$. Moreover, the system performance of the RIS-aided CF system improves more significantly than that of the conventional CF system. In particular, when the number of elements $N$ is sufficiently large, for example, compared with $N=64$, $N=100$ only offers a marginal performance gain which reveals that it is not cost-effective to continuously increase the number of RIS elements in the CF system to achieve further system performance improvement.

Figure~\ref{Figure_3} shows the average SE per UE as a function of the number of AP $M$ and AP antennas $L$ with different $N$. It is clear that increasing the number of AP antennas $L$ can improve the system performance as the increased number of spatial degrees of freedom facilitates more efficient beamforming. Moreover, it is interesting to find that when $N = 64$, the average SE gain of increasing the numbers of AP antennas is much smaller than that $N = 4$ due to channel hardening. This reveals that the number of AP antennas and RIS elements should be determined carefully to realize a cost-effective system.

Figure~\ref{Figure_4} shows the CDF of average SE per UE under the spatial correlations of AP antennas and RIS elements. We consider ${d_{V}}={d_{H}} = {1 \mathord{\left/
 {\vphantom {1 8}} \right.
 \kern-\nulldelimiterspace} 8}\lambda ,{1 \mathord{\left/
 {\vphantom {1 4}} \right.
 \kern-\nulldelimiterspace} 4}\lambda ,{1 \mathord{\left/
 {\vphantom {1 2}} \right.
 \kern-\nulldelimiterspace} 2}\lambda $ denotes the spatial correlation of RIS elements. Let ${\sigma _\varphi } = {5^ \circ }$ and ${{\bf{R}}_{mk}} = \beta _{mk}^{{\rm{NLoS}}}{{\bf{I}}_N}$ represent strong spatial correlation and uncorrelated scenarios of AP antennas, respectively. Note that the existence of correlation has a negative impact on the system performance, especially the correlation in RIS elements which lead to poor passive beamforming at the RIS. Moreover, when ${d_{V} = {1 \mathord{\left/
 {\vphantom {1 8}} \right.
 \kern-\nulldelimiterspace} 8}\lambda}$, the $95\% $-likely performance is the best which is related to the characteristics of sinc function. Note that different from the conclusion which AP antenna spatial correlation is beneficial in \cite{wang2020uplink}, the RIS elements correlation in the aggregated channel affects the AP antennas correlation, resulting in both negative impacts.

Figure~\ref{Figure_5} shows the average SE per UE as a function of the number of APs $M$ for different $N$ for the L-MMSE and the MR combining based on the MMSE estimation. For the L-MMSE combining scheme, the average SE can obtain $10.69\%$ and $7.15\%$ of gain on that of the MR at $M=80$ for $N= 64$ and $N=16$, respectively. It is clear that the performance gap between L-MMSE and MR combining becomes larger with the increase of $N$ since L-MMSE combining can fully exploit all the elements on RIS to suppress interference. Yet, the performance gain of the L-MMSE comes at the expense of involved computation as indicated in (11).

\section{Conclusions}\label{se:conclusion}
In this paper, we study the uplink SE of a spatially correlated RIS-aided CF mMIMO system over spatially correlated channels with multi-antenna APs. For the MMSE channel estimation, we analyzed the uplink SE for the APs adopting the MR/L-MMSE combining and obtained the closed-form SE expressions for the MR combining. It is clear that increasing the number of RIS elements and AP antennas is beneficial to the system performance. Moreover, the spatial correlations of RIS elements and AP antennas both impose a negative impact on the uplink SE and the elements with half-wavelength spacing at the RIS obtain the best performance. Finally, the L-MMSE combining performs better than the MR combining with a large number of RIS elements. In future work, we will consider the power control and the beamforming design to enable the implementation of RIS-aided CF mMIMO networks.

\begin{appendices}

\section{Proof of Theorem 1}
This appendix calculates the covariance matrix in (5). To start with, we express
\begin{align}
&\mathbb{E}\!\!\left\{\! {{ {{{\bf{\tilde o}}}_{mk}}{{{\bf{\tilde o}}}^{H}_{mk}} }} \!\right\} = {{\bf{R}}_{mk}} + {\bf{\bar H}}_m^H{\bf{\Phi}} {{{\bf{\tilde R}}}_k}{{\bf{\Phi}} ^H}{{{\bf{\bar H}}}_m} \notag\\
 &+\!\! \underbrace {\mathbb{E}\!\left\{\! {{\bf{\tilde H}}_m^H{\bf{\Phi}} {{{\bf{\bar z}}}_k}{\bf{\bar z}}_k^H{{\bf{\Phi}} ^H}{{{\bf{\tilde H}}}_m}} \!\right\}}_{{\bf{Q}}^1_{mk}} \!\!+\! \underbrace {\mathbb{E}\!\left\{\! {{\bf{\tilde H}}_m^H{\bf{\Phi}} {{{\bf{\tilde z}}}_k}{\bf{\tilde z}}_k^H{{\bf{\Phi }} ^H}{{{\bf{\tilde H}}}_m}} \!\right\}}_{{\bf{Q}}^2_{mk}}.
\end{align}
To calculate ${{\bf{Q}}^1_{mk}}$, we let ${{\bf{B}}_k} \!\!=\!\! {\bf{\Phi }}{{{\bf{\bar z}}}_k}{\bf{\bar z}}_k^H{{\bf{\Phi }}^H}$, \!${{\bf{Q}}^1_{mk}}$ is derived as
\vspace{-0.4cm}
\begin{align}
{{\bf{Q}}^1_{mk}} = \mathbb{E}\left\{ {\begin{array}{*{20}{c}}
{{\bf{\tilde H}}_{1m}^H{{\bf{B}}_k}{{{\bf{\tilde H}}}_{1m}}}& \!\!\cdots \!\!&{{\bf{\tilde H}}_{1m}^H{{\bf{B}}_k}{{{\bf{\tilde H}}}_{Lm}}}\\
 \!\!\vdots \!\!&\!\! \ddots \!\!& \!\!\vdots \!\!\\
{{\bf{\tilde H}}_{Lm}^H{{\bf{B}}_k}{{{\bf{\tilde H}}}_{1m}}}& \!\!\cdots \!\!&{{\bf{\tilde H}}_{Lm}^H{{\bf{B}}_k}{{{\bf{\tilde H}}}_{Lm}}}
\end{array}} \right\}.
\end{align}
For each element in ${{\bf{Q}}^1_{mk}}$, we can obtain $\mathbb{E}\!\left\{ \!{{\bf{\tilde H}}_{lm}^H{{\bf{B}}_k}{{{\bf{\tilde H}}}_{l'm}}} \!\right\} \!{\buildrel (a) \over =} {\rm{tr}}\!\!\left( \!\!{{{\bf{B}}_k}{{\left[\! {{{{\bf{\tilde R}}}_m}} \!\right]}_{\left( {lN - N + 1 \sim lN,l'N - N + 1 \sim l'N} \right)}}} \!\right)$, where (a) follows by applying the trace of product property ${\rm{tr}}\left( {{\bf{XY}}} \right) = {\rm{tr}}\left( {{\bf{YX}}} \right)$ for some given size-matched matrices ${\bf{X}}$ and ${\bf{Y}}$.
We calculate $ll'$-th element ${\left[ {{\bf{Q}}_{mk}^2} \right]_{ll'}} = {\rm{tr}}\left( {{\bf{\Phi }}{{{\bf{\tilde R}}}_k}{{\bf{\Phi }}^H}{{\left[ {{{{\bf{\tilde R}}}_m}} \right]}_{\left( {lN - N + 1 \sim lN,l'N - N + 1 \sim l'N} \right)}}} \right)$ by applying the same method as above.

\section{Proof of Theorem 2}
The expectations in (14) are calculated here. We begin with the molecular term as
\begin{align}
&\mathbb{E}\left\{ {{{\bf{u}}_{kk}}} \right\} = \mathbb{E}\left\{ {{{\left[ {{\bf{\hat o}}_{1k}^H{{{\bf{\hat o}}}_{1k}}, \cdots ,{\bf{\hat o}}_{Mk}^H{{{\bf{\hat o}}}_{Mk}}} \right]}^T}} \right\}\notag\\
& =\!\! {\left[ {{\rm{tr}}\left( {{\hat p_k}{\tau _p}{{\bf{\Omega }}_{1k}}} \right) \!\!+\!\! {{\left\| {{{{\bf{\bar o}}}_{1k}}} \right\|}^2}\!,\! \cdots \!,\!{\rm{tr}}\left( {{\hat p_k}{\tau _p}{{\bf{\Omega }}_{Mk}}} \right) \!\!+\!\! {{\left\| {{{{\bf{\bar o}}}_{Mk}}} \right\|}^2}} \right]^T}\!\!.
\end{align}
Similarly, we compute the noise term as
\begin{align}
&{\bf{a}}_k^H{{\bf{D}}_k}{{\bf{a}}_k} = {\bf{a}}_k^H{\rm{diag}}\left( {\mathbb{E}\left\{ {{{\left\| {{{{\bf{\hat o}}}_{1k}}} \right\|}^2}} \right\}, \cdots ,\mathbb{E}\left\{ {{{\left\| {{{{\bf{\hat o}}}_{Mk}}} \right\|}^2}} \right\}} \right){{\bf{a}}_k}\notag\\
& \!=\!\! {\bf{a}}_k^H\!{\rm{diag}}\!\left(\! {{\rm{tr}}\!\left( {{\hat p_k}{\tau _p}{{\bf{\Omega }}_{1k}}} \!\right) \!\!+\!\! {{\left\| {{{{\bf{\bar o}}}_{1k}}} \right\|}^2}\!\!, \!\cdots \!\!,{\rm{tr}}\!\left( {{\hat p_k}{\tau _p}{{\bf{\Omega }}_{Mk}}} \right) \!\!+\!\! {{\left\| {{{{\bf{\bar o}}}_{Mk}}} \right\|}^2}} \!\right)\!{{\bf{a}}_k}\notag\\
& = {\bf{A}}_k^H{\rm{diag}}\left( {{z_{1k}}, \cdots ,{z_{Mk}}} \right){{\bf{A}}_k}.
\end{align}

The interference term in the denominator of (14) is
\begin{align}
&\mathbb{E}\left\{ {{{\left| {\sum\limits_{m = 1}^M {a_{mk}^{}{\bf{\hat o}}_{mk}^H{{\bf{o}}_{mi}}} } \right|}^2}} \right\}\notag\\
&= \sum\limits_{m = 1}^M {\sum\limits_{n = 1}^M {{a_{mk}}a_{nk}^ * } } \mathbb{E}\left\{ {{{\left( {{\bf{\hat o}}_{mk}^H{{\bf{o}}_{mi}}} \right)}^H}\left( {{\bf{\hat o}}_{nk}^H{{\bf{o}}_{ni}}} \right)} \right\},
\end{align}
where $\mathbb{E}\left\{ {{{\left( {{\bf{\hat o}}_{mk}^H{{\bf{o}}_{mi}}} \right)}^H}\left( {{\bf{\hat o}}_{nk}^H{{\bf{o}}_{ni}}} \right)} \right\}$ is computed for all possible APs and UEs combinations. We utilize the independence of channel estimation at different APs. When $m \ne n,i \notin {{\cal P}_k}$, we obtain $\mathbb{E}\left\{ {{{\left( {{\bf{\hat o}}_{mk}^H{{\bf{o}}_{mi}}} \right)}^H}\left( {{\bf{\hat o}}_{nk}^H{{\bf{o}}_{ni}}} \right)} \right\} = 0$. For $m \ne n,i \in {{\cal P}_k}\backslash \left\{ k \right\}$, we derive $\mathbb{E}\left\{ {{{\left( {{\bf{\hat o}}_{mk}^H{{\bf{o}}_{mi}}} \right)}^H}\left( {{\bf{\hat o}}_{nk}^H{{\bf{o}}_{ni}}} \right)} \right\} = \mathbb{E}\left\{ {{\bf{\hat o}}_{mi}^H{{{\bf{\hat o}}}_{mk}}} \right\}\left\{ {{\bf{\hat o}}_{nk}^H{{{\bf{\hat o}}}_{ni}}} \right\}$, where
\begin{align}
\mathbb{E}\left\{ {{\bf{\hat o}}_{mi}^H{{{\bf{\hat o}}}_{mk}}} \right\} = \sqrt {{{\hat p}_k}{{\hat p}_i}} {\tau _p}{\rm{tr}}\left( {{\bf{R}}_{mk}^o{\bf{\Psi }}_{mk}^{ - 1}{\bf{R}}_{mi}^o} \right),
\end{align}
since $\mathbb{E}\left\{ {{{{\bf{\bar o}}}^{H}_{mk}}{{{\bf{\bar o}}}_{mi}}{e^{ - j{\varphi _{mk}}}}{e^{j{\varphi _{mi}}}}} \right\} = 0$ and $\mathbb{E}\left\{ {\sqrt {{\hat p_k}{\tau _p}} \left({\bf{R}}_{mk}^o{\bf{\Psi }}_{mk}^{ - 1}\left( {{\bf{y}}_{mk}^p - {\bf{\bar y}}_{mk}^p} \right)\right)^{H}{{{\bf{\bar o}}}_{mi}}{e^{j{\varphi _{mi}}}}} \right\} = 0$. We repeat the same calculation for AP $n$ and obtain
\begin{align}
&\mathbb{E}\left\{ {{{\left( {{\bf{\hat o}}_{mk}^H{{\bf{o}}_{mi}}} \right)}^H}\left( {{\bf{\hat o}}_{nk}^H{{\bf{o}}_{ni}}} \right)} \right\} \notag\\
&= {{\hat p}_k}{{\hat p}_i}\tau _p^2{\rm{tr}}\left( {{\bf{R}}_{mk}^o{\bf{\Psi }}_{mk}^{ - 1}{\bf{R}}_{mi}^o} \right){\rm{tr}}\left( {{\bf{R}}_{ni}^o{\bf{\Psi }}_{nk}^{ - 1}{\bf{R}}_{nk}^o} \right).
\end{align}
For another case $m \ne n,i = k$, we obtain
\begin{align}
&\mathbb{E}\left\{ {{{\left( {{\bf{\hat o}}_{mk}^H{{\bf{o}}_{mi}}} \right)}^H}\left( {{\bf{\hat o}}_{nk}^H{{\bf{o}}_{ni}}} \right)} \right\} \notag\\
&= \hat p_k^2\tau _p^2{\rm{tr}}\left( {{{\bf{\Omega }}_{mk}}} \right){\rm{tr}}\left( {{{\bf{\Omega }}_{nk}}} \right) + {\rm{tr}}\left( {{{{\bf{\bar o}}}_{mk}}{\bf{\bar o}}_{mk}^H} \right){\rm{tr}}\left( {{{{\bf{\bar o}}}_{nk}}{\bf{\bar o}}_{nk}^H} \right)\notag\\
&+ {{\hat p}_k}{\tau _p}{\rm{tr}}\left( {{{\bf{\Omega }}_{nk}}} \right){\rm{tr}}\left( {{{{\bf{\bar o}}}_{mk}}{\bf{\bar o}}_{mk}^H} \right) + {{\hat p}_k}{\tau _p}{\rm{tr}}\left( {{{\bf{\Omega }}_{mk}}} \right){\rm{tr}}\left( {{{{\bf{\bar o}}}_{nk}}{\bf{\bar o}}_{nk}^H} \right).
\end{align}
Similarly for $m = n,i = k$, we utilize the same method as in \cite{ozdogan2019performance} that yields
\begin{align}
&\mathbb{E}\left\{ {{{\left| {{\bf{\hat o}}_{mk}^H{{\bf{o}}_{mk}}} \right|}^2}} \right\} = \hat p_k^2\tau _p^2{\left| {{\rm{tr}}\left( {{{\bf{\Omega }}_{mk}}} \right)} \right|^2} + {{\hat p}_k}{\tau _p}{\rm{tr}}\left( {{{\bf{\Omega }}_{mk}}{\bf{R}}_{mk}^o} \right)\notag\\
&+ {\bf{\bar o}}_{mk}^H{\bf{R}}_{mk}^o{{{\bf{\bar o}}}_{mk}} + {{\hat p}_k}{\tau _p}{\bf{\bar o}}_{mk}^H{{\bf{\Omega }}_{mk}}{{{\bf{\bar o}}}_{mk}} \notag\\
&+ 2{{\hat p}_k}{\tau _p}{\rm{tr}}\left( {{{\bf{\Omega }}_{mk}}} \right){\bf{\bar o}}_{mk}^H{{{\bf{\bar o}}}_{mk}} + {\rm{tr}}{\left( {{{{\bf{\bar o}}}_{mk}}{\bf{\bar o}}_{mk}^H} \right)^2}.
\end{align}
Then, for the case $m = n,i \notin {{\cal P}_k}$, we obtain
\begin{align}
&\mathbb{E}\left\{ {{{\left| {{\bf{\hat o}}_{mk}^H{{\bf{o}}_{mi}}} \right|}^2}} \right\} = {\hat p_k}{\tau _p}{\rm{tr}}\left( {{\bf{R}}_{mi}^o{{\bf{\Omega }}_{mk}}} \right)\notag\\
&+ {\bf{\bar o}}_{mk}^H{\bf{R}}_{mi}^o{{{\bf{\bar o}}}_{mk}} + {\hat p_k}{\tau _p}{\bf{\bar o}}_{mi}^H{{\bf{\Omega }}_{mk}}{{{\bf{\bar o}}}_{mi}} + {\left| {{\bf{\bar o}}_{mk}^H{{{\bf{\bar o}}}_{mi}}} \right|^2}.
\end{align}
For $m = n,i \in {{\cal P}_k}\backslash \left\{ k \right\}$, we obtain
\begin{align}
&\mathbb{E}\left\{ {{{\left| {{\bf{\hat o}}_{mk}^H{{\bf{o}}_{mi}}} \right|}^2}} \right\} = {{\hat p}_k}{\tau _p}{\rm{tr}}\left( {{{\bf{\Omega }}_{mk}}{\bf{R}}_{mi}^o} \right){\rm{ + tr}}\left( {{{{\bf{\bar o}}}_{mk}}{\bf{\bar o}}_{mk}^H{\bf{R}}_{mi}^o} \right)\notag\\
&+ {\left| {{\bf{\bar o}}_{mk}^H{{{\bf{\bar o}}}_{mi}}} \right|^2} + {{\hat p}_k}{{\hat p}_i}\tau _p^2{\left| {{\rm{tr}}\left( {{\bf{R}}_{mk}^o{\bf{\Psi }}_{mk}^{ - 1}{\bf{R}}_{mk}^o} \right)} \right|^2}.
\end{align}
Finally, we can derive the expectation of (14).

\end{appendices}

\bibliographystyle{IEEEtran}
\bibliography{IEEEabrv,Ref}

\end{document}